Title:

# Crack-tip Plasticity and Intrinsic Toughening in Nano-sized Brittle Amorphous Carbon

Dahye Shin[1] and Dongchan Jang[1]*

[1]*Department of Nuclear and Quantum Engineering, Korea Advanced Institute of Science and Technology (KAIST), 291 Daehak-ro, Yuseong-gu, Daejeon 34141, Republic of Korea*

e-mail:

Dahye Shin: sth528@kaist.ac.kr

**Dongchan Jang* (corresponding author):** dongchan.jang@kaist.ac.kr

**Abstract**

Most monolithic brittle materials are vulnerable to the failure by cracks because of a lack of intrinsic toughening mechanisms, such as the plasticity in the vicinity of the crack front. As a result, most of the efforts to mitigate the sudden failure of brittle ceramics have been focused on developing the extrinsic toughening mechanisms that hinder crack propagation behind the tip, such as the fiber bridging. In this work, we experimentally demonstrate that the intrinsic toughening arises even in the brittle monolithic ceramic material such as diamond-like carbon (DLC) when its external dimension reduces down to sub-micron scales. This unique phenomenon owes its origin to the decrease of the crack driving force in the small samples, which in turn enables them to bear high enough stresses to activate the local atomic plasticity. Through nanomechanical tensile and bending experiments, electron energy loss spectroscopy analysis, and finite element method for stress distribution calculation, we confirmed that the local atomic plasticity associated with $sp^3$ to $sp^2$ rehybridization is responsible for the intrinsic toughening.

# 1. Introduction

The recent technological trend toward imparting mechanical deformability to functional devices, such as flexible electronics or bio-implantable gadgets, necessitates the knowledge on the mechanics of functional materials at the scale of an individual component of such systems, typically micron or nano. Ceramics constitute one important class of such functional materials, available properties of which have drastically expanded these days as a result of many new phenomena discovered at the nanometer scale. While there have been a number of studies investigating electronic (Fischetti et al., 2002; Kato et al., 2004; Minot et al., 2003; Mkhoyan et al., 2009; Ruffieux et al., 2012), magnetic (Akbarzadeh et al., 2012; Jiang et al., 2015; Li et al., 2011; Reddy et al., 2012; Saha et al., 2019), plasmonic (Boltasseva and Shalaev, 2015) and optical properties (Aumer et al., 2000; Jacobsen et al., 2006; Kuznetsov et al., 2016; Silambarasu et al., 2018; Yang et al., 2002) of ceramic nanomaterials, the importance of mechanical performances focused on deformation, fracture, and failure has been relatively underestimated. However, to assure the reliable operation and prolonged lifetime of such engineering systems under various loading and straining conditions, it is of great importance to understand the fundamental mechanical deterioration mechanisms, e.g., fracture, that impair the integrity and performance. During the last decade, there have been a lot of works looking into the mechanical behavior of nano-sized materials, many of which mainly focused on the strengthening and toughening of crystalline materials that already have known plasticity mechanisms, such as the dislocation glide (Greer and Nix, 2006; Li et al., 2018; Solanki et al., 2011; Tanaka and Higashida, 2004). However, there is only a little literature so far, which deals with the plasticity-induced toughening effect in the intrinsically brittle amorphous material under given pre-notch conditions. Instead, some studies reported the prevalence of brittle failure down to nanometer scale (Gu et al., 2014; Shimada et al., 2015; Sumigawa et al., 2017).

According to fracture mechanics, the fracture resistance of a material is involved with its intrinsic tolerance to the propagation of cracks, determined as a result of two competing changes with the crack length, i.e., reduction of the total potential energy of the material and loading system (crack driving force, $G$) and increase of free energy in the material due to the creation of new surfaces or any other microstructural changes (fracture resistance, $R$). In the conventional monolithic ceramic materials, most of the mechanical processes capable of hindering crack growth, e.g., a large amount of plasticity, possess much higher activation barrier than the simple cleavage, rendering them to be hardly accessible. As a result, the crack begins to proceed unimpededly once the critical condition has reached, e.g., under the load-controlled tensile loading, and the material fails in a brittle manner with low fracture resistance. Because of this lack of the intrinsic crack obstruction mechanism in ceramics, most of efforts to enhance their fracture toughness have been put into the development of nano-composites that capitalize on the extrinsic mechanism such as the fiber or grain bridging (Chandra et al., 2019; Ohji et al., 1998; Palmero, 2015; Reveron et al., 2017; Sakai et al., 1988; Sternitzke, 1997; Swanson et al., 1987; Wilkerson et al., 2019). On the contrary, the manifestation of the intrinsic toughening process such as the crack tip plasticity has been a long-lasting challenge for monolithic ceramics.

In analogy to many nanomaterials whose properties are distinguished from those of their bulk counterparts, the fracture strength of nano-sized brittle ceramics is also expected to increase with the decrease of the external dimension of the specimen. This phenomenon owes its origin to the reciprocal dependence of the crack driving force ($G$) to the surface-to-volume ratio, thereby decreasing with specimen size. Imagining a self-similar transformation of a cracked body by a particular factor of $\alpha$, i.e., contraction or expansion of all dimensions of the object by the same factor of $\alpha$, we can find that LEFM prescribes $G$ to change to $\alpha G$ if the applied stress remains unchanged (Anderson et al., 2017):

$$G = -\frac{d\Pi}{dA} = \frac{P^2}{2}\frac{dC}{dA} \xrightarrow[\text{transformation}]{\text{self-similar}} \frac{(\alpha^2 P)^2}{2}\frac{d(C/\alpha)}{d(\alpha^2 A)} = \alpha G \qquad \text{Eq. (1)}$$

, where $\Pi$ is the potential energy, $A$ is the crack area, $P$ is the applied load, and $C$ is the compliance of the specimen. This simple relation suggests that the smaller samples ($\alpha < 1$) have less crack driving force under the same applied stress than the larger ones. Accordingly, their fracture strengths should increase to compensate for the reduction of the driving force. This inverse correlation between the sample size and fracture strength in brittle nanomaterials does not only impart high mechanical strength but also offer a new opportunity for a high-barrier physical process to take place by withstanding the high stress field. That is to say, the increased fracture strength allows the formation of the fractionally large plastic zone in the vicinity of the crack tip, in which the local plasticity could occur even for the very brittle ceramics. Under this circumstance, K-dominated zone gradually diminishes with the sample size, and the enlarged plastic zone starts to play important roles in the fracture behavior. In this regard, the size and shape of the plastic zone under the given loading and geometric condition become the key factors in understanding the mechanical properties of brittle nanomaterials.

In this study, we performed the nanoscale uniaxial tensile and fixed-end bending tests to experimentally demonstrate the significance of the crack-tip plasticity and its influence on the mechanical responses of the nano-sized monolithic diamond-like carbon (DLC). From a series of the uniaxial tension experiments with varying the sample size and notch geometry, we observed that the fracture toughness increases with the local stress field near the crack tip, implying the considerable contribution from the plastic work. In addition, in order to investigate the crack-tip plasticity under the stress condition where shear components prevail, we conducted the fixed-end bending fracture tests. Interestingly, the unusual nonlinear response beyond the initial linear elasticity appears in the load-displacement curves in the course of crack propagation, which is likely attributed to the extensive plastic work due to the dominance of shear stresses under bending of the nano-beams. Furthermore, the measurement of local atomic orbital concentration using electron energy loss spectroscopy (EELS) in the scanning transmission electron microscope (STEM) reveals that the $sp^2/sp^3$ number ratio distribution coincides with that of the plastic strain, suggesting that $sp^3$-to-$sp^2$ rehybridization concurrently occurs with local atomic plasticity event.

## 2. Methods

### 2.1 Material Preparation

The DLC film was coated on the Si wafer to 1.2 μm thickness using Plasma Enhanced Chemical Vapor Deposition at low temperature (PECVD, *Nanocoat-1000*) by *K-DLC co.* Hydrogen content of the DLC film was analyzed separately using two independent methods: visible Raman spectroscopy with a wavelength of 514 nm (*ARAMIS*) and hydrogen determinator (ELTRA *ONH-2000*), each of which measures the concentration of between 0.84 and 3.55 at.% and 0.63 at.%, respectively. While the former has relatively large uncertainty due to its indirect comparison to the previous experimental data, both results confirm that our DLC film can be classified into low hydrogen content amorphous carbon (Casiraghi et al., 2005; Robertson, 2002). Non-crystallinity of the film was verified using High-Resolution (HR) Transmission Electron Microscope (TEM) imaging and Selected Area Diffraction (SAD) in *Titan cubed G2 60-300* operating at 200 kV. In addition, the small correlation length of about 10 Å from the Raman spectroscopy further supports the amorphous nature of the DLC coating. Young's modulus was measured by nanoscale flexural tests and nanoindentation with continuous stiffness measurement (CSM) option at the depth where indentation size effect becomes saturated (around 120 nm). Each measurement gave a very similar value of 180 GPa. Detailed information is available in Supplementary Discussion 1.

### 2.2 Fabrication of Nano-specimens for Mechanical testing

Miniaturized test specimens for nanoscale tensile and bending fracture experiments were fabricated horizontally on the DLC film surface along the edge of the wafer piece using focused ion beam (FIB, FEI *Helios 450 F1*) milling, during which rectangular patterns were sequentially used on the top and side faces (see Figure 1(a) and (b)) at the final $Ga^+$ ion beam condition of 30 kV and 33 pA. The gauge section of the dog-bone-shaped tensile specimens was constructed after tilting the incident ion beam by 45 degrees from the plane normal direction of the film. As schematically illustrated in Figures 1(a) and (b), each sample contains 60 nm- or 50 nm-deep pre-notches at the center of its length on the surface, created for tensile and bending fracture tests, respectively. The tensile specimens were fabricated into three different geometries, all with the same length of 1 μm but differing in the width and pre-notch configuration: Double-Edge-Notched-Tension samples with a width of 200 nm (DENT200), 300 nm (DENT300), and 400 nm (DENT400) and Single-Edge-Notched-Tension with a width of 200 nm (SENT200). For bending fracture tests, Fixed-end Single-Edge-Notched-Bending (F-SENB) samples were fabricated with a width of 500 nm and a length of 3 μm with both ends elastically clamped. Dimensions and the representative SEM images of each sample are summarized in Table 1 and Figures 1(c) through (d). Samples for nanoscale flexural tests for elasticity measurement were fabricated in the same way to the bending fracture test specimens but without pre-notch (see supplementary Figure S1(b)).

**2.3 Nanomechanical Experiments**

The *in-situ* nanoscale uniaxial tensile and bending tests were carried out using the *PI-87* picoindenter (*Bruker, Inc*) equipped with a tension grip or wedge tip (see Figure 1(h) and (i)), respectively, inside the Scanning Electron Microscope (SEM, Hitachi *SU5000*) under the electron beam condition of 5 kV and spot size 30. The grip and tip were fabricated using the FIB (FEI *Quanta*) out of the Boron-doped conductive diamond and precisely aligned with samples prior to each test. All tests were performed at a constant strain rate in a range between $1\times10^{-3}$ s$^{-1}$ and $4\times10^{-3}$ s$^{-1}$. The thermal drift rate was maintained to be smaller than 0.05 nm/s, which contributes only less than 1% to the total displacements. Some selected bending samples were intentionally unloaded before they completely broke, and then further thinned to the electron-transparent thickness, on which to conduct the electron energy loss spectroscopy (EELS) analysis near the crack tip. Nine nanoindentation tests were done with a maximum depth of 120 nm (~ 10% of film thickness) at the strain rate of $10^{-3}$/s under CSM mode.

**2.4 Finite-element methods (FEM)**

The stress distribution in the specimen was calculated using the finite element method (FEM) for each of DENT200, DENT300, DENT400, SENT200 and F-SENB with U-shaped slender notches of 10 nm tip radius, assuming the two-dimensional plane strain condition. Only isotropic elasticity was used for the constitutive relationship with Young's modulus and Poisson's ratio of 180 GPa and 0.22 (Cho et al., 1999), respectively. The mesh densities in the body were approximately 18,500 and 6,500 per square micrometer for tensile and bending simulations, respectively, with 8-node biquadratic plane strain quadrilateral (CPE8) element type. Meshes near the crack tip were refined further to have a higher density than the remaining part. For all tensile specimens, the rotational degree of freedom was removed at the top surface, while the bottom end was completely fixed to the rigid body. For F-SENB, both ends were rigidly clamped, and the frictionless hard contact was assumed between the rigid circular indenter (100 nm radius) and the top surface.

The load-displacement curve for F-SENB sample (Figure 5(b)) was generated by performing a series of independent FEM calculations with varying loads and crack lengths. For each run, we fixed the crack length, and continuously monitored the reaction force acted on the rigid indenter while pushing the indenter into the bending specimen until the force satisfies the critical J-integral condition. Critical J-integral was determined by plugging experimentally-determined critical displacement at the initial crack length, 50 nm, i.e. displacement value at mark 1 in Figure 3(d). Once completed, a similar calculation was repeated after increasing the crack length.

**2.5 STEM-EELS and data treatment**

Electron Energy Loss Spectroscopy (EELS) data in the Scanning Transmission Electron Microscope (STEM) mode were collected using *Titan cubed G2 60-300* at KAIST Analysis Center for Research Advancement (KARA) with the operating voltage of 200 kV. The energy resolution was approximately 1 eV as confirmed by the Full Width at Half Maximum (FWHM) of the zero-loss peak, and semi-collection angle (β) was fixed at 100 mrad for all measurements. The pixel dwell time of 0.5 sec and 0.25 eV/channel dispersion were used. Two separate regimes in the EELS spectra were obtained for each specimen, i.e., the low-loss between −10 eV and 100 eV (valence electron excitation) to deconvolute the core-loss spectra and high-loss from 240 eV to 350 eV (core electron excitation). In order to obtain the EELS signals near the growing crack, we prepared the unbroken bending specimen by intentionally unloading before the crack reached the other side and subsequently thinned it down to approximately 70 nm thickness by the FIB. The core-loss spectra were analyzed following the straight path connecting the crack tip and one of the fixed ends to see the fractional change of $sp^2/sp^3$ carbon bond ratio as a function of the location.

Typical K-edge of DLC shows two characteristic peaks, one around 285 eV originated from the 1s→π* excitation (π*-peak) and the other above 290 eV from 1s→σ* excitation (σ*-peak). Given that both $sp^3$ and $sp^2$ bonds contribute to the σ*-peak, but π*-peak solely comes from the $sp^2$, the number fraction of $sp^2$ and $sp^3$ orbitals is related with the π* and σ* peak intensities (Egerton, 2011) following:

$$\frac{I_{\pi*}}{I_{\sigma*}} = \frac{N_\pi}{N_\sigma} \cdot \frac{P_{\pi*}(\beta\Delta)}{P_{\sigma*}(\beta\Delta)} \qquad \text{Eq. (2)}$$

, where $I_{\pi*}$ and $I_{\sigma*}$ are the integrated peak intensities, $P_{\pi*}(\beta\Delta)$ and $P_{\sigma*}(\beta\Delta)$ are partial ionization cross sections depending on the semi-collection angle (β) and the integration range (Δ), and $N_\pi$ and $N_\sigma$ are the numbers of π- or σ-bonded carbon, respectively. Even though both of the instrumental condition (β) and the data processing parameter (Δ) are needed to specify the particular number fraction of each carbon bond, in this work we only display the relative variation with respect to the value at the uttermost location from the crack tip through the following relationship:

$$\frac{\left[{}^{I_{\pi*}}/_{I_{\sigma*}}\right]_{\text{position x}}}{\left[{}^{I_{\pi*}}/_{I_{\sigma*}}\right]_{\text{position ref.}}} = \frac{\left[{}^{N_{\pi}}/_{N_{\sigma}}\right]_{\text{position x}}}{\left[{}^{N_{\pi}}/_{N_{\sigma}}\right]_{\text{position ref.}}}. \qquad \text{Eq. (3)}$$

To determine the peak intensities, we first removed the background signal by extrapolating the pre-edge data across the peak, then deconvoluted the effects of the multiple electron scattering using Fourier-ratio method (Williams and Carter, 1996), and finally applied the so-called 'two-window method' (Bruley et al., 1995), width and location of which are marked in yellow in Figure 3(l). This analysis is sensitive to the choice of the location and width ($\Delta$) of each window, especially that for the $\pi^*$ peak, $\Delta_{\pi^*}$, so, they need to be determined with great care (Bernier et al., 2008; Bruley et al., 1995; Zhang et al., 2016). In this work, we followed the criterion offered by Bruley et al. (Bruley et al., 1995), according to which the energy window is positioned at the center of each peak with the width of 2 eV for $\Delta_{\pi^*}$ and 10 eV for $\Delta_{\sigma^*}$.

## 3. Results

Figure 2 shows the results of nano-tensile fracture tests conducted on four different types of samples, i.e., DENT200, DENT300, DENT400, and SENT200. Fracture always occurred at the pre-notch, as shown in Figure 2(c). Nominal engineering stress-strain curves are presented in Figures 2(d) through (g) for DENT200, DENT300, DENT400, and SENT200, respectively. The nominal stress is obtained by dividing the applied load by the gross area, not by the net area between pre-notches. Fracture strengths of all four samples, determined as the engineering stress at failure, are similar to each other around at ~2 GPa. These values are considerably high, approaching the significant fraction of the theoretical tensile strength of this material, (~E/10 = 18 GPa) (Frenkel, 1926; Gao et al., 2003). These high strengths can be attributed to the enhanced crack stability at nano-sized materials caused by the reduction of the crack driving force as discussed in the Introduction. It also allows the nano-sized materials to withstand the large crack tip stresses, offering an additional possibility for a stress-induced structural change to occur there. Interestingly, even if the DENT200 specimen has the longest total notch in the fractional sense (*w/2a*, see Table 1), it can withstand almost the same far-field tensile stress with the others. This result suggests that DENT200 would have higher fracture resistance than the others.

Figure 3 presents the result of bending experiments on F-SENB specimens. Figure 3(a) through (c) show snapshots taken during F-SENB tests (a) at contact, (b) ~180 nm flexural displacement, and (c) after complete failure. Load-displacement curves measured at the strain rate of $10^{-3}$ sec$^{-1}$ given in Figure 3(d) are composed of the initial linear elastic loading during which no crack growth occurs, followed by the onset of nonlinearity (marked by 1) at which the crack begins growing, and subsequent load drop (marked by 2). Then, the crack grows steadily in a nonlinear manner during which the load gradually increases (between 2 and 3), and the final fracture (marked by 3) occurs at a load of ~250 µN and a displacement of ~200 nm. This nonlinearity, accompanied by the monotonic increase of the load and simultaneous crack growth, appears to be nontrivial for the materials whose behavior is prescribed only by linear elasticity. Moreover, Figure 3(e) shows that the load-displacement relation remains unchanged while changing the strain rate over two orders of magnitude. The absence of strain rate effect on this nonlinearity presumably suggests that the physical process behind it is driven by the stress, but not thermally activated. In order to further investigate any structural changes involved with the strain-hardening-like nonlinearity, we analyzed the electron energy loss spectroscopy (EELS) in the scanning transmission electron microscope (STEM) at three different crack conditions, i.e., two specimens with straight crack up to 50% (Figure 3(g)) and 80% (Figure 3(h)) of the sample width and one with curvy growth path (Figure 3(i)). Figure 3(f) shows the load-displacement curves from the bending experiments conducted to make such partially-cracked specimens. To create the curved crack, we intentionally misoriented the tip to push the sample with an angle of 45º (see Supplementary Figure S5) and unloaded after the significant amount of the nonlinear deformation proceeded. Noticeably, the mechanical response from the slant-loading (black in Figure 3(f)) is the smoothest compared with the others and does not show any discontinuous load drops. Bright-field and high-resolution TEM images presented in Figure 3(j) demonstrates that the stress-induced crystallization did not take place in the vicinity of the crack. The red arrow on the SEM image of an F-SENB sample after the unloading test in Figure 3(k) indicates the scanning path of the electron beam to collect EELS spectra in the STEM mode. In Figure 3(l), typical C-K edge in the EELS spectra of DLC are given with two peaks, $\pi^*$-peak near 285 eV and $\sigma^*$-peak above 290 eV. Yellow windows indicate the integration range of each peak needed to apply the two-window method mentioned in section 2.5.

## 4. Discussions

Commonly-used physical measure to quantify the fracture resistance of the material is the fracture toughness. Being an intrinsic property of material but dictated by the stress states near the singularity, the fracture toughness has conventionally been assessed with the well-defined experimental setup that excludes the influences from the geometric conditions, e.g., with perfectly sharp crack tip and sufficiently large specimen. However, the intrinsic smallness of nanomaterials renders those conventional conditions to be unavailable. For example, the curvature of pre-notch in this work, ~ 10 nm, would be small enough to be considered ideally sharp in the typical bulk samples, but the round-tip effect must be taken into account for nano-sized samples because it is not ignorable any more compared with the notch length of ~60 nm. Therefore, in order to reliably attain fracture toughness from the nano-tensile experiments, we need to carefully analyze any effects from the non-standard sample geometry, including the crack-tip roundness and finite sample dimensions. It is well-known that under the uniaxial tensile loading condition the stress concentration factor (SCF) at the end of a slender notch with a finite curvature is related with the tensile stress intensity factor $K_I$ via (Pilkey, 2008):

$$K_I = (\text{SCF} - 1) \cdot \sigma_{\text{far}}\sqrt{\pi\rho}/2 \qquad \text{Eq. (4)}$$

, where $\rho$ is the radius of curvature of notch tip and $\sigma_{\text{far}}$ is the uniaxial tensile stress far from the notch. Plugging the experimentally measured fracture strength for $\sigma_{\text{far}}$ and numerically calculated SCF using the same geometric parameters with the experiments into Eq. 4, we obtained fracture toughness $K_{\text{IC}}$ of 1.01 MPa√m, 0.80 MPa√m, 0.81 MPa√m, and 0.82 MPa√m for DENT200, DENT300, DENT400 and SENT200, respectively (see Table 2). The numerically calculated SCFs of DENT specimens are comparable with the literature values (Pilkey, 2008; Tada et al., 2000), further validating the accuracy of our calculation. DENT200 samples have a higher fracture toughness than the others by about 25%. With all geometric factors decoupled, the increase of fracture toughness in DENT200 sample is likely attributed to the occurrence of intrinsic toughening mechanism. In other words, the local stress field near the notch reaches high enough magnitude to activate the atomic-level plastic shearing in the amorphous samples, which in turn gives the plastic work effect to the fracture toughness. To validate this possibility, we conducted further calculations to estimate the amount of plastic energy of each sample.

Figure 4(a) shows the distribution of Tresca stress (twice of the maximum shear stress) calculated using FEM under the uniaxial tensile stresses equivalent with experiments, i.e., 2050 MPa, 1930 MPa, 2040 MPa and 2050 MPa for DENT200, DENT300, DENT400, and SENT200, respectively (see Table 2). The regions colored in red in the insets indicate the area where the maximum shear stresses exceed the theoretical shear strength $\tau_{\text{th}}$ of DLC, which is estimated to be $\mu/30 = 2.47$ GPa, with $\mu$ the shear modulus (Anderson et al., 2017; Glezer and Shurygina, 2017; Simmons et al., 1970; Stachurski, 2011, 2015), and therefore represent the plastic zones of our samples at fracture. It can be clearly seen that DENT200 has a larger plastic zone than the others because the stress fields from each pre-notch overlap. Once the plasticity arises in the course of the crack growing, it makes additional contribution to the total critical energy release rate ($G_R$) by the amount of $dT_R/dA$ (Anderson, 2017), where $T_R$ is plastic shear energy in a given plastic zone ($A_p$) per unit thickness and $A$ is the crack surface area. Then, $dT_R/dA$ term during the incremental crack growth from the initial length, $a_0$, by $da$ can be estimated as following (details in Supplementary Discussion 2):

$$\left.\frac{dT_R}{dA}\right|_{a=a_0} = \tau_{\text{th}} \gamma_p \left[\frac{dA_p}{da}\right]_{a=a_0}. \qquad \text{Eq. (5)}$$

, where $\tau_{\text{th}}$ and $\gamma_p$ are theoretical shear strength ($\mu/30$) and associated plastic shear strain (~1/30) (Johnson and Samwer, 2005) in the amorphous solids, respectively. Because the critical energy release rate ($G_R$) and fracture toughness ($K_{\text{IC}}$) are correlated with each other via $G_R = K_{\text{IC}}^2/E$ (Irwin, 1957), the difference in the measured fracture toughness ($K_{IC}$) between the samples, e.g., between DENT 200 and DENT 400, becomes

$$\frac{1}{E}\left[ K_{IC}^2 \Big|_{\text{DENT200}} - K_{IC}^2 \Big|_{\text{DENT400}} \right] = \tau_{th}\gamma_p \left( \left[\frac{dA_p}{da}\right]_{a=a_0}^{\text{DENT200}} - \left[\frac{dA_p}{da}\right]_{a=a_0}^{\text{DENT400}} \right) \quad \text{Eq. (6)}$$

, where we assumed that the contribution from the intrinsic brittleness, such as the surface energy of DLC, remains unchanged regardless of the specimen geometries. Substituting 2.47 GPa, 1/30, and 14.7 nm for $\tau_{th}$, $\gamma_p$, and $\Delta(dA_p/da)|_{a_0}$, respectively, the right-hand-side (RHS) of Eq. (6) becomes ~1.21 J/m². Here, we obtained the $(dA_p/da)|_{a_0}$ term from a series of FEM simulations with slightly changing the crack length while maintaining the external load constant (see Supplementary Discussion 3 for the detailed procedures). On the other hand, the left-hand-side (LHS) of Eq. (6) is immediately available from the experimentally measured fracture toughness and Young's modulus of 180 GPa (see Table 2 and Supplementary Figure S1(c)), and equals to ~2.10 J/m². It should be noted that our FEM calculation only serves as the first order approximation with simplified elastic constitutive relation and crudely-estimated plasticity threshold. Therefore the discrepancy between the calculation (RHS) and the experimental measurement (LHS) is likely due to the overestimation of the stress magnitude within the plastic zone as well as the underestimation of the plastic zone size itself. Nonetheless, our model reveals that the effect of small-scale yielding cannot be ignored even in the brittle materials when the sample size is reduced to nanoscales. Instead, it can contribute to the increase of the fracture toughness as much as ~25% for DENT200 specimen. Figure 4(b) graphically shows this positive correlation between the plastic shear energy and the fracture toughness.

Unlike the uniaxial tensile loading condition, it is not straightforward to quantitatively measure the fracture toughness from the fixed-end bending. However, we can still qualitatively compare the crack-tip plastic zone of DENT200 and F-SENB at the onset of the fracture using the same FEM analysis. As shown in Figure 5(a), the Tresca stress distribution of F-SENB at the moment of the initial crack growth, i.e., corresponding to the point marked by 1 in Figure 3(d), spreads out wider, and the equivalent plastic zone (marked red in the inset) is larger than those in the tensile specimens. At least in the qualitative sense, this relative enlargement of plastic zone due to the shear-dominant stress state of bending suggests that F-SENB sample has much enhanced fracture resistance than the other tensile specimens. For example, the load-displacement curves in Figure 5(b) compare the mechanical responses of our sample with the one obtained using FEM simulations. As opposed that the purely-elastic numerical computation predicts the significant load drop upon initiation of crack growth, the actual F-SENB nanobeam sample underwent the gradual load increase even during crack propagation. While the former is in good agreement with the conventional observation, the improved mechanical stability of the latter can be attributed to the large plastic energy at the nano-scale.

The continuum-level stress analysis demonstrates that plasticity-induced toughening phenomenon can occur even in the highly brittle materials once their extrinsic dimensions reduce to the nanoscale, but the atomistic process that accommodates such plastic work in the amorphous carbon needs to be specified. One possible mechanism is the atomic scale plastic shearing event during which a few atoms shift one another to form a small cluster that carries the local atomic-level plastic strain, i.e., phenomenologically similar idea with the shear transformation zone (STZ) in metallic glasses (Argon, 1979), though the covalent bonding in DLC may cause different characteristics. As illustrated in Figures 6 (a) and (b), the continuum-level plastic strain is determined by the population of such clusters, each of which is represented by a red dot. In the high-stress region, such as the crack tip vicinity, the high density of the clusters results in the large plastic strain, while they distribute sparsely in the low-stress place and the plastic strain is negligibly small there. In conventional brittle materials, as shown in Figure 6(a), the plastic zone with a high density of clusters is significantly smaller than the K-dominated zone. Even though there may exist a few of them because of the structural heterogeneity, their population is too low to mitigate the brittle fracture in the K-dominated zone. However, at the nanoscale (Figure 6(b)), the decrease in the crack driving force enables the nano-samples to withstand high applied stress, which in turn creates the larger high-stress region near the crack tip than their bulk counterpart. Consequently, the large number of shear clusters within the zone leads to the high plastic strain, which contributes to the increase of the plastic shear energy, $T_R$ in Eq (5), and therefore the fracture toughness. Under this circumstance, the spatial distribution of $sp^2/sp^3$ ratio can serve as an experimental tracer for the local plastic deformation in DLC materials because the carbon orbital rehybridization from $sp^3$ to $sp^2$ may concurrently occur with this atomic shearing process (Liu and Meletis, 1997; Voevodin et al., 1996) – or we can equivalently consider the orbital rehybridization as one of the specific atomic shearing mechanisms in DLC. In most amorphous carbon materials, both $sp^2$ and $sp^3$ atomic structures co-exist (Kowalczyk et al., 2012; Marks et al., 1996; Robertson and O'reilly, 1987; Theye and Paret, 2002) and the externally-applied high shear stresses assist transitions between them (Bouchet et al., 2015; Chen et al., 2015; Gao et al., 2002; Kunze et al., 2014; Li et al., 2019; Ma et al., 2014; Pastewka et al., 2011; Romero et al., 2014; Sanchez-Lopez et al., 2003). This dependence of rehybridization on the shear stress magnitude enables us to experimentally assess the distribution of local plastic deformation of DLC by using $sp^2/sp^3$ ratio as an indicator for the amount of local plastic strain.

In Figure 7(a), the normalized $I_{\pi*}/I_{\sigma*}$ ratios from EELS spectra as a function of the distance from the crack tip (following the red arrow in Figure 3(k)) are presented for three types of F-SENB samples (refer to Figures 3(f) through (i)). Given that the normalized $I_{\pi*}/I_{\sigma*}$ scales with the fractional concentration of $sp^2$ orbitals, higher this value means more $sp^3$-to-$sp^2$ transition occurs. Figure 7(b) displays the variation of the maximum shear stress along the same path with EELS scanning, calculated using FEM at the loading condition equal to the ones used to prepare EELS specimen. Here, we can clearly see that the normalized $I_{\pi*}/I_{\sigma*}$ ratios qualitatively follow the same trend with the shear stress distributions. This observation confirms that the atomic plastic shearing events actually take place, assisted by the shear stress. Moreover, $I_{\pi*}/I_{\sigma*}$ ratio from the curvy cracked sample (black) is the highest, i.e., almost 28% higher than the lowest value of the 50%-straightly-cracked one. Overall, $I_{\pi*}/I_{\sigma*}$ ratio is the smallest for 50%-straightly-cracked sample (blue) followed by 80%-straightly-cracked (green) and curvy-cracked ones (black) in the increasing order. This trend is opposed to the amount of load drops in Figure 3(f). Likewise, the curved-crack sample shows the largest plastic shear energy in the plastic zone, followed by 80% and 50% cracked ones in the decreasing order as in Figure 7(c). This positive correlation between the concentration of $sp^2$ bonds, i.e., the amount of plastic strain, and the plastic shear energy further supports that the occurrence of local plasticity is responsible for enhancing the mechanical performance, such as the increase of fracture toughness, in the DLC nano-materials. In addition, it has been reported that $sp^3$ to $sp^2$ transition in amorphous carbon is a mechanically driven athermal process rather than caused by the elevation of temperature (Kunze et al., 2014; Ma et al., 2014; Romero et al., 2014), which explains the strain-rate independency we observed in F-SENB tests (Figure 3(e)).

## 4. Conclusions

In this work, the fracture resistance of nano-scaled DLC was measured from various tensile fracture tests. Mode I fracture toughness of Double-Edge-Notch-Tension (DENT) specimen with width of 200 nm is larger by ~25% (1 MPa√m) than the others (0.8 MPa√m). This enhancement of the fracture resistance is attributed to the decrease in the crack driving force due to the size reduction to the nanometer orders and consequent activation of local plasticity caused by the large notch tip stress fields. In addition, through Fixed-End Single-Edge-Notched-Bending (F-SENB) tests and EELS characterization, we further verified the positive correlation between the shear stress distribution and $sp^2$ bond concentration, i.e., ~28 % higher $sp^2/sp^3$ fraction in the shear-maximized condition. Based on those observations, we conclude that fracture resistance of nano-sized DLC is enhanced through the crack-tip shear stress-induced $sp^3$ to $sp^2$ local structure transformation, enabled by the increased crack stability at nanoscale. These findings offer a new opportunity to tailor the fracture behavior by appropriately designing the crack configuration and sample geometry in the brittle nanomaterials.

## 4. Acknowledgements

The authors acknowledge financial support from National Research Foundation of Korea (NRF- NRF-2019M2D2A1A02038972).

**Figure & Table Captions**

**Figure 1** Schematic illustrations of fabrication procedures of (a) tensile fracture specimens, (b) fixed-end bending fracture specimens using focused-ion-beam (FIB). SEM images of fabricated samples: (c) DENT200, (d) DENT300, (e) DENT400, (f) SENT200, (g) F-SENB, and tips: (h) tension grip and (i) wedge tip for bending. Scale bars, 500 nm (c-g), 2 μm (h), 5 μm (i).

**Figure 2** SEM images (a) after contact, (b) right before the failure, (c) after fracture, and nominal engineering stress-strain curves for (d) DENT200, (e) DENT300, (f) DENT400 and (g) SENT200 from nano-tensile fracture tests. Scale bars, 1 μm (a-c).

**Figure 3** SEM images from F-SENB tests (a) at contact, (b) during crack growing, and (c) after failure. Load-displacement curves for F-SENB tests (d) until fracture, (e) under different strain rate conditions, and (f) unloaded before fracture resulting in different crack geometries. TEM images of (g) 50%, (h) 80% straightly grown crack and (i) crack under slanted loading. (j) TEM image near the curved grown crack, (k) SEM image of TEM sample prepared by FIB with EELS data acquisition path (red arrow), and (l) obtained EELS spectra with energy windows for two-window method. Scale bars, 500 nm (a-c, k), 100 nm (g-i), 10 nm (j).

**Figure 4** (a) Tresca stress distribution obtained by FEM at failure and related crack tip plastic zone (red area) of tensile specimens. (b) Correlation between shear energy and fracture toughness for each specimen type. Scale bar, 20 nm (a).

**Figure 5** (a) Tresca stress distribution and plastic zone (red area) at the moment of crack initiation of F-SENB specimen. (b) Comparison of experimental and FEM load-displacement curves. Scale bar, 40 nm (a).

**Figure 6** Schematic illustration of the relationship between continuum stress fields and the local atomic plasticity which accommodated by $sp^3$ to $sp^2$ carbon transition (a) in a bulk material and (b) a nano-sized material.

**Figure 7** (a) Normalized $I_{\pi^*}/I_{\sigma^*}$ ratio along the distance from the crack tip. (b) The maximum shear stress plot along the path that EELS spectra obtained, with Tresca distributions and (c) shear energy in the plastic zone from FEM at unloading moment. Scale bar, 500 nm (b).

**Table 1** Summary of sample labeling and dimensions.

**Table 2** Measured fracture strength from nanoscale tensile fracture tests, stress concentration factors from FEM, and obtained fracture toughness values.

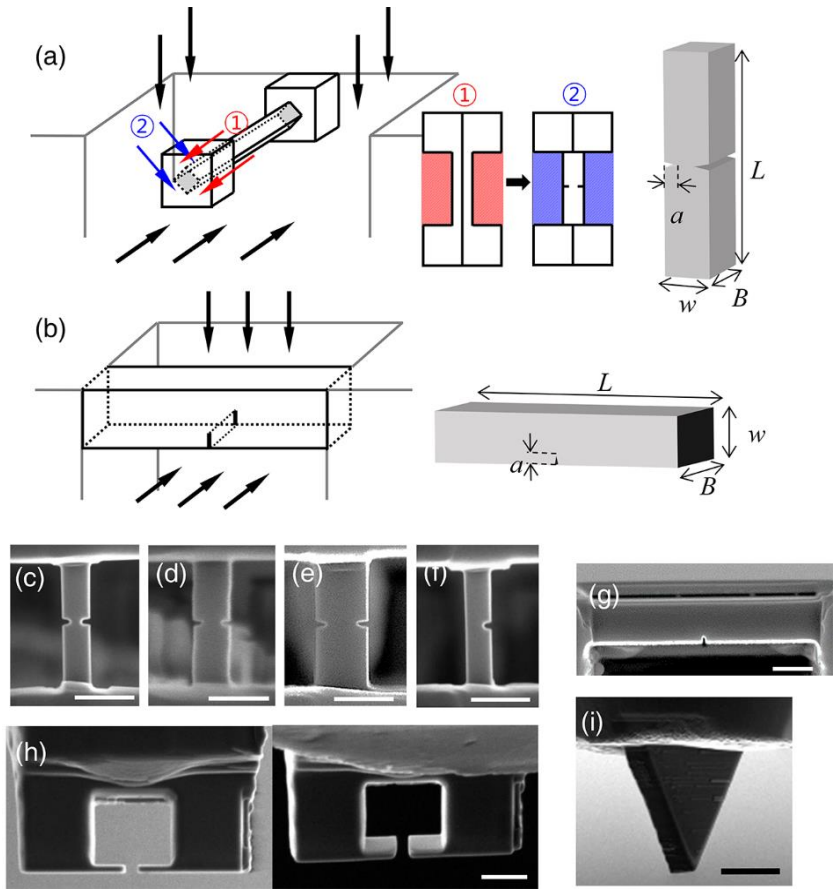

**Figure 1.** Schematic illustrations of fabrication procedures of (a) tensile fracture specimens, (b) fixed-end bending fracture specimens using focused-ion-beam (FIB). SEM images of fabricated samples: (c) DENT200, (d) DENT300, (e) DENT400, (f) SENT200, (g) F-SENB, and tips: (h) tension grip and (i) wedge tip for bending. Scale bars, 500 nm (c-g), 2 μm (h), 5 μm (i).

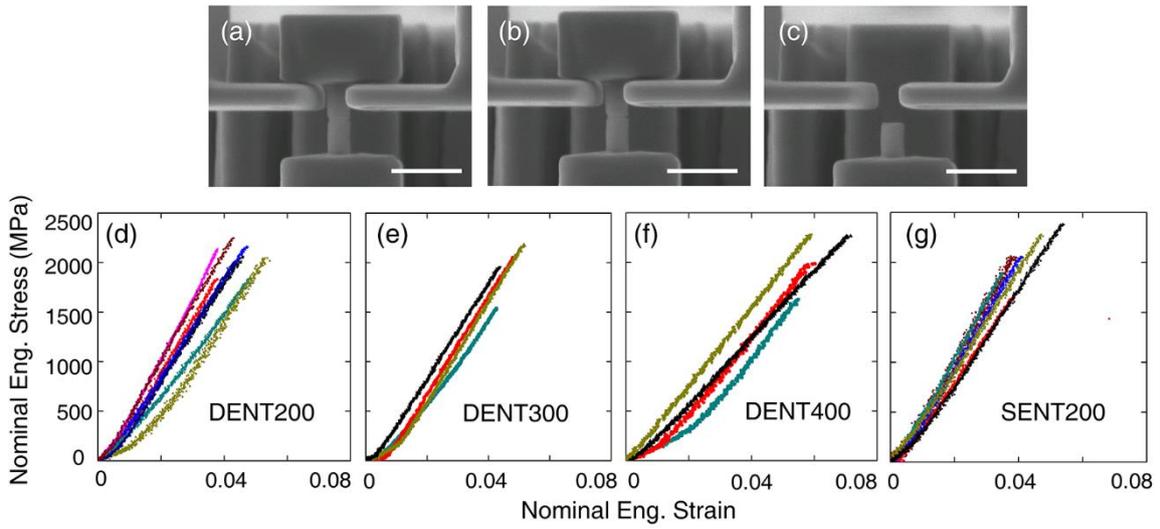

**Figure 2.** SEM images (a) after contact, (b) right before the failure, (c) after fracture, and nominal engineering stress-strain curves for (d) DENT200, (e) DENT300, (f) DENT400 and (g) SENT200 from nano-tensile fracture tests. Scale bars, 1 μm (a-c).

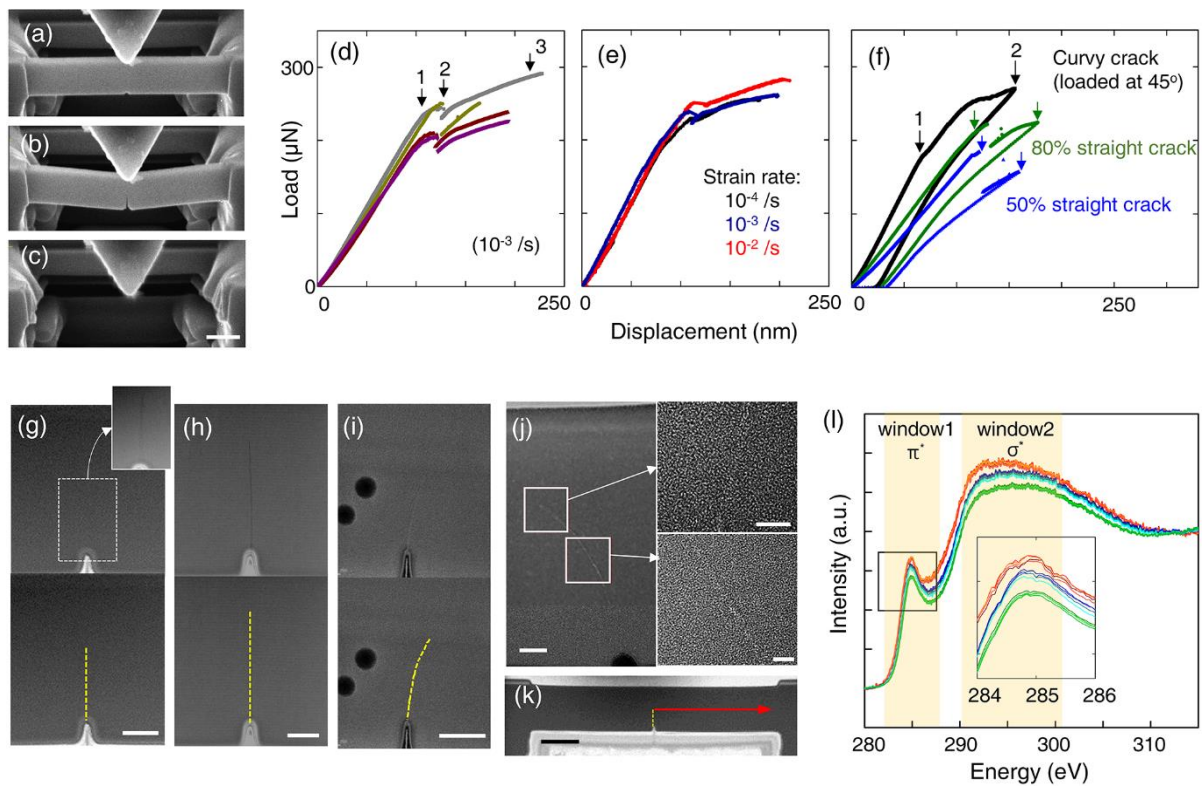

**Figure 3.** SEM images from F-SENB tests (a) at contact, (b) during crack growing, and (c) after failure. Load-displacement curves for F-SENB tests (d) until fracture, (e) under different strain rate conditions, and (f) unloaded before fracture resulting in different crack geometries. TEM images of (g) 50%, (h) 80% straightly grown crack and (i) crack under slanted loading. (j) TEM image near the curved grown crack, (k) SEM image of TEM sample prepared by FIB with EELS data acquisition path (red arrow), and (l) obtained EELS spectra with energy windows for two-window method. Scale bars, 500 nm (a-c, k), 100 nm (g-i), 10 nm (j).

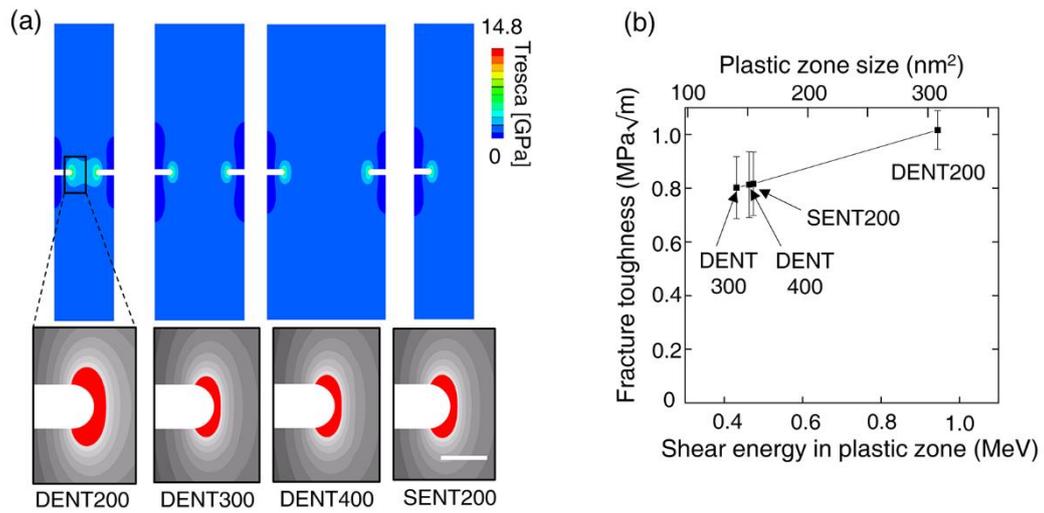

**Figure 4.** (a) Tresca stress distribution obtained by FEM at failure and related crack tip plastic zone (red area) of tensile specimens. (b) Correlation between shear energy and fracture toughness for each specimen type. Scale bar, 20 nm (a).

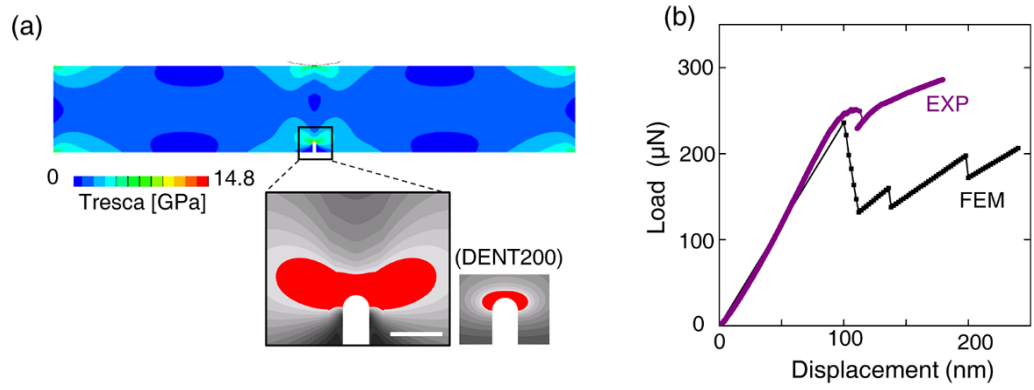

**Figure 5.** (a) Tresca stress distribution and plastic zone (red area) at the moment of crack initiation of F-SENB specimen. (b) Comparison of experimental and FEM load-displacement curves. Scale bar, 40 nm (a).

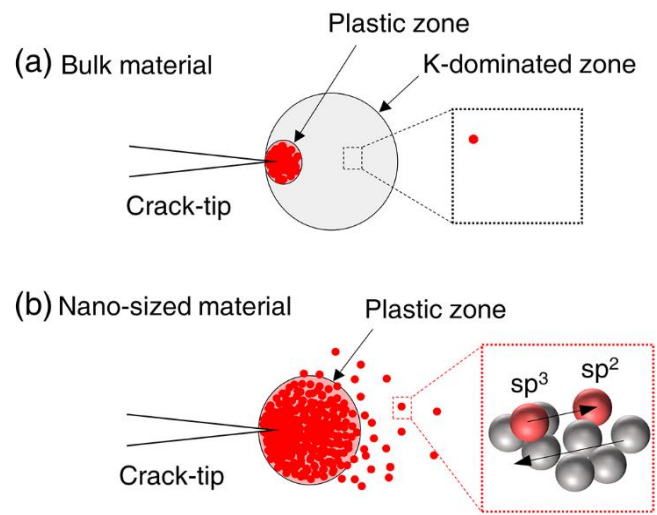

**Figure 6**. Schematic illustration of the relationship between continuum stress fields and the local atomic plasticity which accommodated by $sp^3$ to $sp^2$ carbon transition (a) in a bulk material and (b) a nano-sized material.

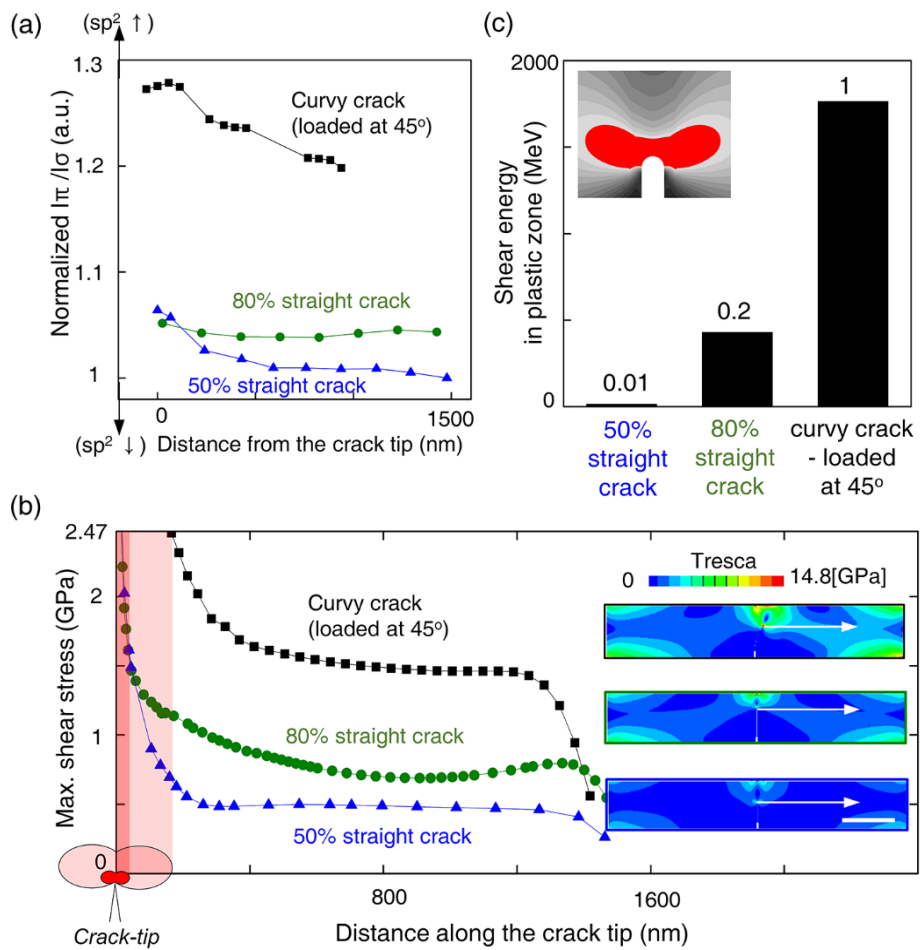

**Figure 7.** (a) Normalized $I_{\pi^*}/I_{\sigma^*}$ ratio along the distance from the crack tip. (b) The maximum shear stress plot along the path that EELS spectra obtained, with Tresca distributions and (c) shear energy in the plastic zone from FEM at unloading moment. Scale bar, 500 nm (b).

| Loading Mode | Tensile | | | | Bending |
|---|---|---|---|---|---|
| Specimen Name | DENT200 | DENT300 | DENT400 | SENT200 | F-SENB |
| Width, $w$ (nm) | 200 | 300 | 400 | 200 | 500 |
| Thickness, $B$ (nm) | 200 | 300 | 400 | 200 | 500 |
| Length, $L$ (µm) | 1 | 1 | 1 | 1 | 3 |
| Pre-notch, $a$ (nm) | ~ 60 | ~ 60 | ~ 60 | ~ 60 | ~ 50 |
| Notch tip radius of curvature, $\rho$ (nm) | ~10 | ~10 | ~10 | ~10 | ~10 |

**Table 1.** Summary of sample labeling and dimensions.

|  | DENT200 | DENT300 | DENT400 | SENT200 |
|---|---|---|---|---|
| Avg. fracture strength ($\sigma_f$) [MPa] | 2048.7 ±135.78 | 1926.4 ±240.19 | 2040.4 ±281.93 | 2049.6 ±265.21 |
| Stress Concentration Factor (SCF) | 6.6 | 5.7 | 5.5 | 5.5 |
| Avg. fracture toughness ($K_{IC}$) [MPa√m] | 1.0168 ±0.06738 | 0.80238 ±0.10004 | 0.81370 ±0.10583 | 0.81697 ±0.10577 |

**Table 2.** Measured fracture strength from nanoscale tensile fracture tests, stress concentration factors from FEM, and obtained fracture toughness values.

# Supplementary Information

## Crack-tip Plasticity and Intrinsic Toughening in Nano-sized Brittle Amorphous Carbon

Dahye Shin[1] and Dongchan Jang[1]*

[1]*Department of Nuclear and Quantum Engineering, Korea Advanced Institute of Science and Technology (KAIST), 291 Daehak-ro, Yuseong-gu, Daejeon 34141, Republic of Korea*

e-mail:

Dahye Shin: sth528@kaist.ac.kr

**Dongchan Jang* (corresponding author):** dongchan.jang@kaist.ac.kr

**This file includes:**
Supplementary Discussions 1 – 4
Supplementary Figures S1 – S5
References for Supplementary Information.

# Supplementary Discussion 1.

# Results of Mechanical Characterization of DLC film and Analysis of visible Raman Spectra

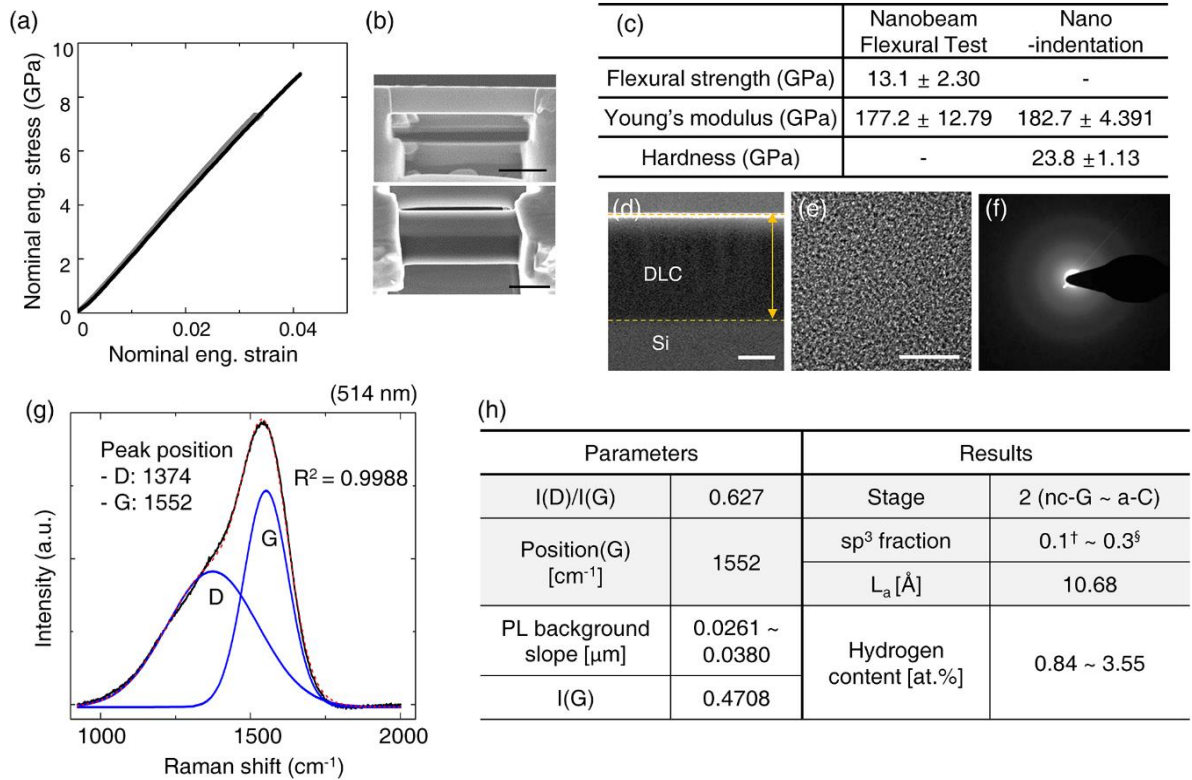

**Supplementary Figure S1.** (a) Nominal engineering stress-strain curves obtained from nanobeam flexural tests of un-notched samples ($L$ = 3 μm, $B$, $w$ = 500 nm) and (b) SEM image of the as-fabricated nano-beam. (c) Mechanical properties measured from nano-beam flexural tests and nanoindentations. (d) SEM image of DLC coating on Si wafer. (e) HR-TEM image and (f) SAD pattern of DLC coating. (g) Visible Raman spectrum with Gaussian fitting, and (h) summary of the fitting parameters and the results. Scale bars, 1 μm (b), 500 nm (d), 10 nm (e). [†Robertson, 2002; §Tamor et al., 1994]

Figure S1(g) displays the visible Raman (514 nm) spectroscopy data fitted to the Gaussian functions (Casiraghi et al., 2005). Similarly to most carbon materials, they contains typical D and G peaks (Tuinstra and Koenig, 1970) at 1374 cm$^{-1}$ and 1552 cm$^{-1}$, respectively. In order to extract the basic information on the DLC sample used in this work, such as sp$^3$ content, the degree of amorphousness, and hydrogen content, from the Raman spectra, we used the phenomenological three-stage model (Ferrari and Robertson, 2000). This model conceptually classifies the structural states of the carbon materials into three stages according to the relative prevalence of structural features observed in the Raman spectroscopy: perfect graphite to nano-crystalline graphite (stage 1), nano-crystalline graphite to amorphous carbon (stage 2), and amorphous carbon to tetrahedral carbon (ta-C) or diamond (stage 3). Considering the intensity ratio of D and G peaks, and the position of G-peak, the DLC film in this work turns out to belong to the 'stage 2', materials in which typically contain relatively low concentration of sp$^3$-bonded structures around 10% - 20%. Quantitative comparison of our data with those in the previous studies reveals that the sample in this work has about 10% (Robertson, 2002) or 30% (Tamor and Vassell, 1994) of sp$^3$ bonding fraction. In-plane cluster size of sp$^2$ carbons, called conjugated length, $L_a$, can also be calculated from the intensity ratio, I(D)/I(G), for stage 2 carbons as (Ferrari and Robertson, 2000)

$$\frac{I(D)}{I(G)} = C'(\lambda) \cdot L_a^2 \text{ . ----- eq (S1.1)}$$

, where I(D) and I(G) are the intensities of D- and G-peak, respectively, determined as peak height after Gaussian fitting, and C'(λ) is a proportionality constant for different wavelength used to obtain Raman spectra, which is 0.0055 Å$^{-2}$ for 514 nm (Ferrari and Robertson, 2000). Relatively small conjugated length of ~ 10 Å was calculated, which implies that only short-range-ordered clusters exist in our DLC coating. Hydrogen content was estimated to be around 0.84 - 3.55 at. % from the slope of photoluminescence background, $m$, normalized by the intensity of G peak, $m$/I(G), as in the following equation (Casiraghi et al., 2005),

$$H\,[\text{at.}\,\%] = 21.7 + 16.6 \log\left\{\frac{m}{I(G)}[\mu m]\right\} \text{ . ----- eq (S1.2)}$$

This amount of hydrogen is low enough to assume our DLC to be Hydrogen free during Raman data analysis (Casiraghi et al., 2005). The hydrogen content independently measured using the element determinator is 0.63 at%, which further confirms the low hydrogen content of our DLC film.

**Supplementary Discussion 2.**

**Derivation of Shear Energy Equation inside the Plastic Zone**

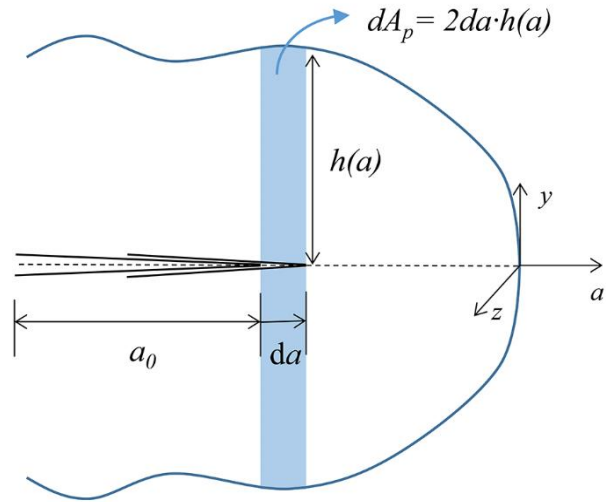

**Supplementary Figure S2.** Scheme of crack-tip plastic zone when crack propagates from $a_0$ to $a_0 +$ d$a$ with specimen thickness, $B$.

If crack surface area, $A$ is assumed to be $Ba$, where $B$ is thickness of the sample and $a$ is crack length, then d$T_R$/d$A$ becomes,

$$\left.\frac{dT_R}{dA}\right|_{a=a_0} = \left.\frac{dT_R}{Bda}\right|_{a=a_0}. \quad \text{--- eq. (S2.1)}$$

, where d$T_R$/d$a$ can be obtained by assuming the symmetric zone shape to y-axis and no energy density variance in z-axis (see figure S.2), as following (Anderson, 2017):

$$\left.\frac{dT_R}{da}\right|_{a=a_0} = 2B \int_0^{h(a)} \left[\int_0^{\varepsilon_{ij}} \sigma_{ij}\, d\varepsilon_{ij}\right] dy. \quad \text{--- eq. (S2.2)}$$

If we suppose the shear stress inside the plastic zone is uniform at the theoretical shear strength, $\tau_{th}$, Eq. S2.2 for d$T_R$/d$a$ can be further simplified to,

$$\left.\frac{dT_R}{da}\right|_{a=a_0} = 2B \int_0^{\frac{dA_p}{2da}} [\tau_{th}\, \gamma_p]\, dy = \tau_{th}\, \gamma_p B \left[\frac{dA_p}{da}\right]_{a=a_0}. \quad \text{--- eq. (S2.3)}$$

, where $\gamma_p$ is theoretical shear strain, associated with $\tau_{th}$ (Johnson and Samwer, 2005). Plugging Eq. S2.3 to Eq. S2.1, d$T_R$/d$A$, then we can finally obtain the expression in Eq (5)

$$\left.\frac{dT_R}{dA}\right|_{a=a_0} = \left.\frac{dT_R}{Bda}\right|_{a=a_0} = \tau_{th}\, \gamma_p \left[\frac{dA_p}{da}\right]_{a=a_0}.$$

**Supplementary Discussion 3.**

**Estimation of the value of <dA_p/da> at a_0**

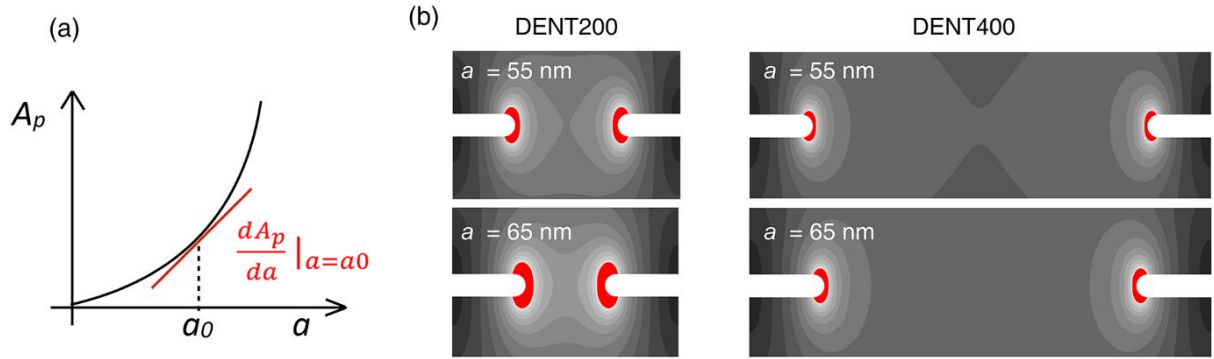

**Supplementary Figure S3.** (a) Schematic plot for the conceptual meaning of d$A_p$/d$a$ at $a_0$. (b) Comparison of change in the plastic zone (red area) at the constant fracture load when varying the notch size ±5 nm from the initial pre-notch, 50 nm, calculated from FEM.

In order to estimate the value of d$A_p$/d$a$ at $a_0$, the slope in $a$- $A_p$ domain at $a_0$ (see S3(a)) was numerically estimated as following. The evolution of the plastic zone area (d$A_p$) has been monitored while changing the notch length by ±5 nm from the initial size, $a_0$ = 50 nm, using FEM under the fracture load for each sample geometry. Then the obtained 'd$A_p$' was divided by delta notch size (d$a$), 10 nm. Finally the differences between each specimen, $\Delta$(d$A_p$/d$a$)|$a_0$ for as ~14.7 nm and used in Eq (5).

**Supplementary Discussion 4.**

**Advantages of Bending Mode and Crack Stability of the F-SENB Specimen**

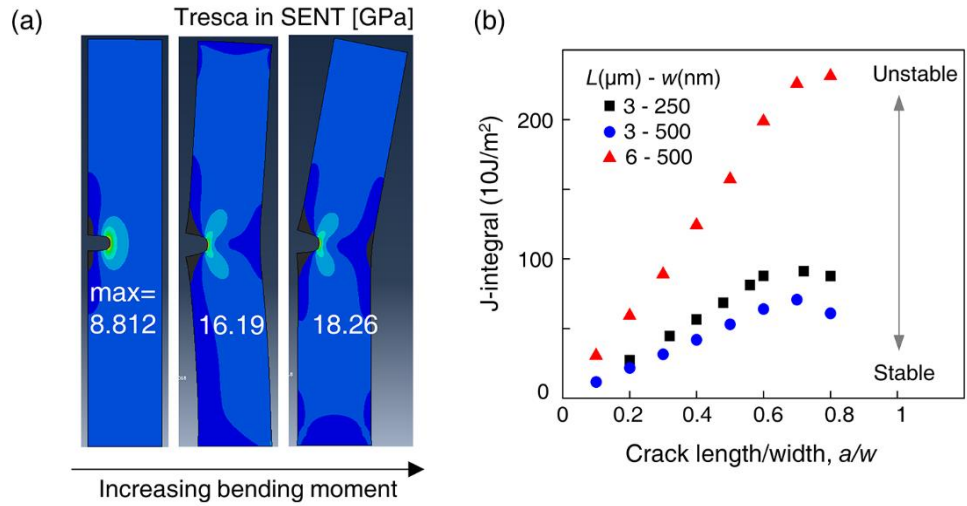

**Supplementary Figure S4.** (a) Maximum Tresca values as virtually increasing the rotational motion on the top surface in SENT from FEM calculations, and (b) J-integral versus crack size plot for different F-SENB geometries with fixed $B = 500$ nm under constant load condition.

In S4(a), we can see that the maximum Tresca stress in the vicinity of the notch increases as the rotational motion of the top surface of the SENT sample is allowed. This observation sheds light on determining the suitable loading mode to investigate shear-induced rehybridization. S4(b) shows a plot of the J-integral versus relative crack size to width ($a/w$) for the F-SENB samples with different aspect ratios and sizes. Unlike the load-controlled tensile test where crack driving force keeps increasing until failure, J-integral of F-SENB reaches the maximum before it completely breaks and decreases thereafter. Furthermore, 3-500 sample has the lowest J-integral values, followed by 3-250 and 6-500 in the ascending order, suggesting that 3-500 has the least crack driving force. Based on this analysis, we chose the F-SENB sample with 3:500 ($L(\mu m)$-$w(nm)$) aspect ratio, with which to perform the near-crack-tip EELS analysis.

The data in S4(b) was obtained by preforming a series of independent FEM simulations while gradually increasing the notch depth from 50 nm under the fixed load. J-integral values were calculated along closed paths around pre-defined crack tip and crack extension direction. At crack growth initiation when $a/w = 0.1$, the rigid indenter was placed at the center of the top surface and deflected it by 100 nm, 200 nm and 400 nm for 3-500, 3-250, and 6-500 samples, respectively to ensure the same beam curvature between the samples of different dimensions. Then, the reaction forces back to the circular rigid indenter were calculated and used as the constant load values for the subsequent calculations with longer cracks.

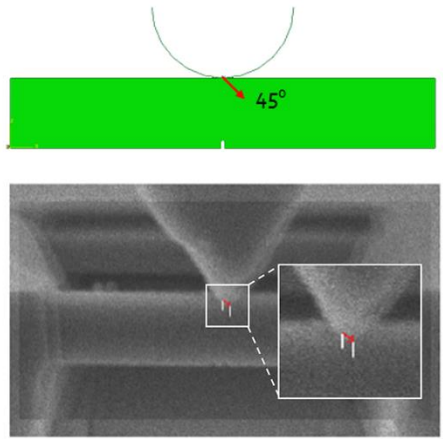

**Supplementary Figure S5.** Scheme of the shear-enhanced system by slant loading and the actual indentation path indicated by a red arrow in an overlap of snapshots from the experiments at mark 1 and 2 in Figure 3(f).